\documentclass[aip,jcp,floatfix]{revtex4-2}

\usepackage[utf8]{inputenc}
\usepackage[range-units=single]{siunitx}
\usepackage{graphicx}
\usepackage{bm}
\usepackage{amsmath}
\usepackage{amssymb}
\usepackage{mathtools}
\usepackage{csquotes}
\usepackage{hyperref}
\usepackage[ISO]{diffcoeff}
\usepackage[dvipsnames]{xcolor}
\usepackage[version=4,arrows=pgf]{mhchem}

\DeclareMathOperator{\cov}{cov}

\DeclarePairedDelimiter{\pqty}{(}{)}
\DeclarePairedDelimiter{\sqty}{[}{]}
\DeclarePairedDelimiter{\bqty}{\lbrace}{\rbrace}
\DeclarePairedDelimiter{\aqty}{\langle}{\rangle}

\DeclarePairedDelimiter{\norm}{\|}{\|}

\DeclareSIUnit\atom{atom}
\DeclareSIUnit\calorie{cal}

\newcommand{\supar}[1]{\ensuremath{^{\pqty*{#1}}}}

\begin{document}

\title{Deep Ensembles vs. Committees for Uncertainty Estimation in Neural-Network Force Fields: Comparison and Application to Active Learning}

\author{Jes\'us Carrete}
\email{jesus.carrete.montana@tuwien.ac.at}
\affiliation{Institute of Materials Chemistry, TU Wien, A-1060 Vienna, Austria}
\author{Hadri\'an Montes-Campos}
\affiliation{Grupo de Nanomateriais, Fot\'onica e Materia Branda, Departamento de F\'{\i}sica de Part\'{\i}culas, Universidade de Santiago de Compostela, E-15782 Santiago de Compostela, Spain}
\affiliation{CIQUP, Institute of Molecular Sciences (IMS)—Departamento de Química e Bioquímica, Faculdade de Ciências da Universidade do Porto, Rua Campo Alegre, 4169-007 Porto, Portugal}
\author{Ralf Wanzenb\"ock}
\affiliation{Institute of Materials Chemistry, TU Wien, A-1060 Vienna, Austria}
\author{Esther Heid}
\affiliation{Institute of Materials Chemistry, TU Wien, A-1060 Vienna, Austria}
\author{Georg K. H. Madsen}
\affiliation{Institute of Materials Chemistry, TU Wien, A-1060 Vienna, Austria}

\date{\today}

\begin{abstract}
A reliable uncertainty estimator is a key ingredient in the successful use of machine-learning force fields for predictive calculations. Important considerations are correlation with error, overhead during training and inference, and efficient workflows to systematically improve the force field. However, in the case of neural-network force fields, simple committees are often the only option considered due to their easy implementation. Here we present a generalization of the deep-ensemble design, based on multiheaded neural networks and a heteroscedastic loss, that can efficiently deal with uncertainties in both the energy and the forces. We compare uncertainty metrics based on deep ensembles, committees and bootstrap-aggregation ensembles using data for an ionic liquid and a perovskite surface. We demonstrate an adversarial approach to active learning to efficiently and progressively refine the force fields. That active learning workflow is realistically possible thanks to exceptionally fast training enabled by residual learning and a nonlinear learned optimizer.
\end{abstract}

\pacs{}

\maketitle

\section{Introduction}
Advanced classification and regression methods are propelling a revolution in computational chemistry and materials science.\cite{von2020retrospective,goh2017deep,mater2019deep} 
Recent advances in statistical modeling open the door to tackling problems of previously prohibitive levels of complexity, such as the discovery or inverse design of compounds or materials in the vastness of chemical space,\cite{von2020exploring,liu2020machine} the prediction of diverse chemical properties of gas-phase or condensed systems,\cite{wu2018moleculenet,chen2018rise,ekins2016next,butler2018machine} or the exploration of chemical reactions.\cite{coley2018machine}

From the point of view of these disciplines, the conceptual and computational pipeline from physical principles to measurable properties of technological importance typically begins at the electronic level. In particular, the scalability of methods based on density functional theory (DFT) makes them a frequent choice to model the behavior of electrons.\cite{kohn1965self} However, this stage often still represents the bottleneck of the calculation. Therefore, although machine learning (ML) has the potential to jump directly from atomistic structure -- or even from less detailed descriptions like a chemical formula -- to quantitative or categorical results, bypassing just the electronic calculation is a very attractive option to develop general, detailed and transferable ML-powered models.\cite{von2020retrospective}

That is precisely the goal of ML force fields (FFs), regression models that take sets of atomic coordinates and atom types as inputs and provide a quantitative picture of the potential energy hypersurface.\cite{unke2021machine} Within the limits of the Born-Oppenheimer approximation, they allow practitioners to retain the extensive apparatus developed for use with semiempirical potentials, including but not limited to molecular dynamics (MD) and Monte Carlo (MC) methods. Critically, however, they do so without sacrificing precision and accuracy with respect to a direct ab-initio approach.

A major downside of using a regression model to calculate energies and forces is the loss of a direct connection to the underlying physics, and therefore the lack of an aprioristic framework to assess the results. Although DFT itself involves uncontrolled approximations that make the quality of its results for a new system uncertain,\cite{goddft} the much more flexible functional forms used in ML can easily lead to vastly larger errors or completely unphysical results for certain values of the input variables. While the most extreme of such mispredictions are likely to become apparent, the effect of subtler errors can become significantly more pernicious as they accumulate along an MD or MC trajectory and potentially invalidate large amounts of data.\cite{fu2022forces}

Another important connection between the training data and the error estimate appears in the context of active learning approaches. Refining a model by expanding the training set, whether it is to describe new phases, to improve its precision within known regions of configuration space or to strike any other specific balance between exploration and exploitation, requires being able to gauge the level of certainty about each prediction.
Ideally, the uncertainty in a prediction made by a ML model correlates with the error, because that allows for active learning approaches to iteratively improve the prediction capabilities of a model by identifying regions of high uncertainty (and therefore presumably high error) and retraining the model after including more data points from these regions.

Words like error and uncertainty are sometimes used rather loosely and the meanings attributed to them in different sources can diverge significantly. In this article we adopt a frame of reference aligned with the recommendations contained in the metrology guides issued by standards bodies like the ISO.\cite{JCGMVIM3,JCGMGUM} Those guidelines have historically moved from the so-called \emph{error approach} or \emph{true value approach} to the modern \emph{uncertainty approach}, where the objective of a measurement is to assign a range of reasonable values to the measurand. We take the energies and forces calculated with DFT as our conventional values, also used as the ground truth for the ML models. Error is defined as the deviation of a predicted value from the conventional value, and uncertainty as a non-negative parameter characterizing the dispersion of the values being attributed to a prediction.

The question of how to evaluate the error and uncertainty in energies and forces is key to enabling robust and reliable MLFF-based workflows that are readily amenable to automation. It is common to find this issue framed in terms of interpolation vs. extrapolation with respect to the training data; that is, however, too simplistic a picture, since regression models in high dimensionality will extrapolate with overwhelming probability.\cite{extrapolation} Broadly speaking, it is not unreasonable to assume that similarity to the data included in the training set will play a role in the quality of the predictions.\cite{hirschfeld2020uncertainty} However, in a setting where sophisticated nonlinear transformations of the input constitute the whole purpose of the calculation, exactly how and even in which space that similarity is to be measured is a question better tackled with help from the model itself.

Uncertainty about the predictions of a general ML model can have two conceptually divergent origins, each of which requires a targeted treatment. The most common approaches try to relate the uncertainty to the error associated with the model's ability to learn from the given data, thus measuring the \emph{epistemic} error inherent to the model itself. Those approaches differ in their complexity, cost, quality of calibration, and ease of implementation on top of existing architectures. In the case of neural networks (NNs), well known strategies include distance-based and nonparametric methods\cite{janet2019quantitative,tran2020methods} and Bayesian NNs where weights and biases have associated probability distributions, as well as the broad family of ensemble-based estimators where randomness is introduced into the process by training several models, the ensemble members. The nature of the ensemble is determined by how those models differ; among the recipes that have been explored we can cite bootstrap aggregation (with members trained on different subsets of the data),\cite{Efron1979} dropout (where a fraction of the connections between neurons is randomly removed),\cite{gal2016dropout} or committees (where randomness is limited to the initialisation of the weights and biases).\cite{politis1994large}

Other approaches such as  mean variance estimation,\cite{nix1994estimating} or evidential learning\cite{soleimany2021evidential} furthermore include sources of error stemming from the data itself, the \emph{aleatoric} error. They assume an error distribution that is a function of the input descriptors (usually Gaussian noise with differing magnitudes) and train the model to learn this distribution along with the prediction of the target. However, by the nature of the training process these models not only fit the input error distribution, but also discount training data points that do not fit well within the current model prediction,\cite{nix1994estimating} so that in practice the putative estimate of the aleatoric uncertainty also contains epistemic contributions. If there is little or mostly uniform noise on the target, the epistemic contribution may even dominate. In general, the sources of error in a model vastly differ depending on the chosen representation of a chemical structure, the architecture of a model, and the size and quality of the training set, and largely influence the ability of uncertainty metrics to correlate with the actual error.\cite{heid2023characterizing} Too little or low quality training data or too restrictive a model usually cause uncertainty metrics to fail, i.e. to not correlate with the actual error anymore.\cite{heid2023characterizing}

Among the most flexible strategies to account for both epistemic and aleatoric components of the uncertainty is another ensemble-based approach, namely deep ensembles.\cite{lakshminarayanan2017simple} Each NN making up a deep ensemble has an additional output that is interpreted as an estimate of the variance of the main quantity being predicted. This is complemented by the use of a heteroscedastic loss, i.e., a loss function where different input points carry different statistical weights that are functions of their variance. Depending on the circumstances, the contribution of each point to the loss can be minimized by pushing its prediction closer to the ground truth or by increasing the estimate of its uncertainty, but there is no risk of a drift towards predicting a high uncertainty for all points because that would lead to a very high loss. Thus, a successful minimization of the loss requires achieving a balance between raw accuracy and predicted uncertainty, which enables simultaneous training of both outputs.

The approaches described above are usually deployed for ML models directly predicting a target quantity, such as electronic properties of a compound. In contrast, MLFFs need to predict not only the energies but also the forces (i.e. the derivatives of the energies) of each atom in a compound or material to allow for an application within MD simulations. Critically, the set of energies and forces have to fulfill the fundamental symmetries of mechanics leading to the conservation of linear and angular momentum. Different MLFF designs have been described, ranging from the preprocessing of the inputs into symmetry-compliant descriptors\cite{behler2007generalized} or the progressive building of atomic environments starting from an overcomplete set of interatomic distances\cite{unke2018reactive} to explicitly equivariant operations on the inputs leading to intermediate and output quantities with well defined tensor ranks.\cite{batzner20223,liao2022equiformer} Likewise, different MLFF architectures have been proposed, including kernel methods\cite{chmiela2017machine,Bartok_PRB13} and a plethora of different NNs like traditional multilayer perceptrons (MLPs),\cite{behler2007generalized} graph neural networks\cite{gilmer2017neural,schutt2017schnet,gasteiger2020directional} and variations on the transformer model.\cite{liao2022equiformer}
Previous uncertainty estimates in MLFFs have been strongly dependent on the underlying design. For instance, for Bayesian inference they have been obtained directly from the covariance matrix. For NNs, the choice has so far been mostly driven by the ease of implementation. Thus, the preferred approaches have been ensemble-based estimators like bootstrap-aggregation ensembles and committees.\cite{Artrith_PRB12,Smith_JCP18,Zhang_JCP18,Musil_JCTC19,Schran_JCP20,MontesCampos_JCIM22}

The complexity of some of their functional forms emphasizes the importance of modularity in MLFFs, and in particular the inconvenience of triggering large implementation efforts subsequent to small changes in the architecture only to obtain consistent derivatives. Automatic differentiation offers a convenient and efficient route to satisfy this need in the context of MLFFs implemented on top of modern frameworks.\cite{MontesCampos_JCIM22} Moreover, it opens the door to Sobolev training,\cite{sobolev} i.e., to the inclusion of derivatives of the target in the loss to force the model to match them. This is known to lead to faster and more accurate results for general function approximation through NNs\cite{NNs_with_derivatives} and is particularly relevant for NNFFs given the limited information content of the total potential energy.\cite{MontesCampos_JCIM22} Considering this, an automatically differentiable NNFF constitutes a good foundation to implement and evaluate more advanced uncertainty estimators.

This paper introduces a strategy to use a deep ensemble of descriptor-based NNFFs. The resulting ensemble can provide uncertainties in the forces on atoms, and not only in the total energy. This requires a generalization of the original deep-ensemble template.\cite{lakshminarayanan2017simple} Although forces are partial derivatives of the energy, we show that the uncertainty in the forces cannot be easily obtained through the application of differential operators to the uncertainty in the energy. We then analyze whether deep ensembles actually provide, in practice, a more useful uncertainty estimate. Training ensembles is more demanding because the probability of a suboptimal training run can be significant enough that one model in an ensemble might not converge to an acceptable minimum of the loss landscape. We achieve a much more stable training process by introducing a deep residual network (ResNet) architecture.\cite{ResNet} We improve its convergence further through the use of the recently released nonlinear learned optimizer VeLO.\cite{VeLO}

The resulting training is orders of magnitude faster, which, in combination with the uncertainty estimates, makes active learning cycles possible. To test our models and workflow we employ data for two systems: first, the ionic liquid ethylammonium nitrate (EAN), where we show how MD runs that result in configurations with a high uncertainty and error can be automatically detected and repaired; second, the surface reconstructions of the perovskite \ce{SrTiO3}, for which, starting from pure bulk structure data, we demonstrate how the MLFF can be iteratively improved by a simple active learning approach based on the maximization of an adversarial loss\cite{Bombarelli_adversarial} to select highly uncertain but physically plausible configurations.

\section{Methods}

\subsection{Revised NeuralIL architecture}

\begin{figure*}
    \centering
    \includegraphics[width=\textwidth]{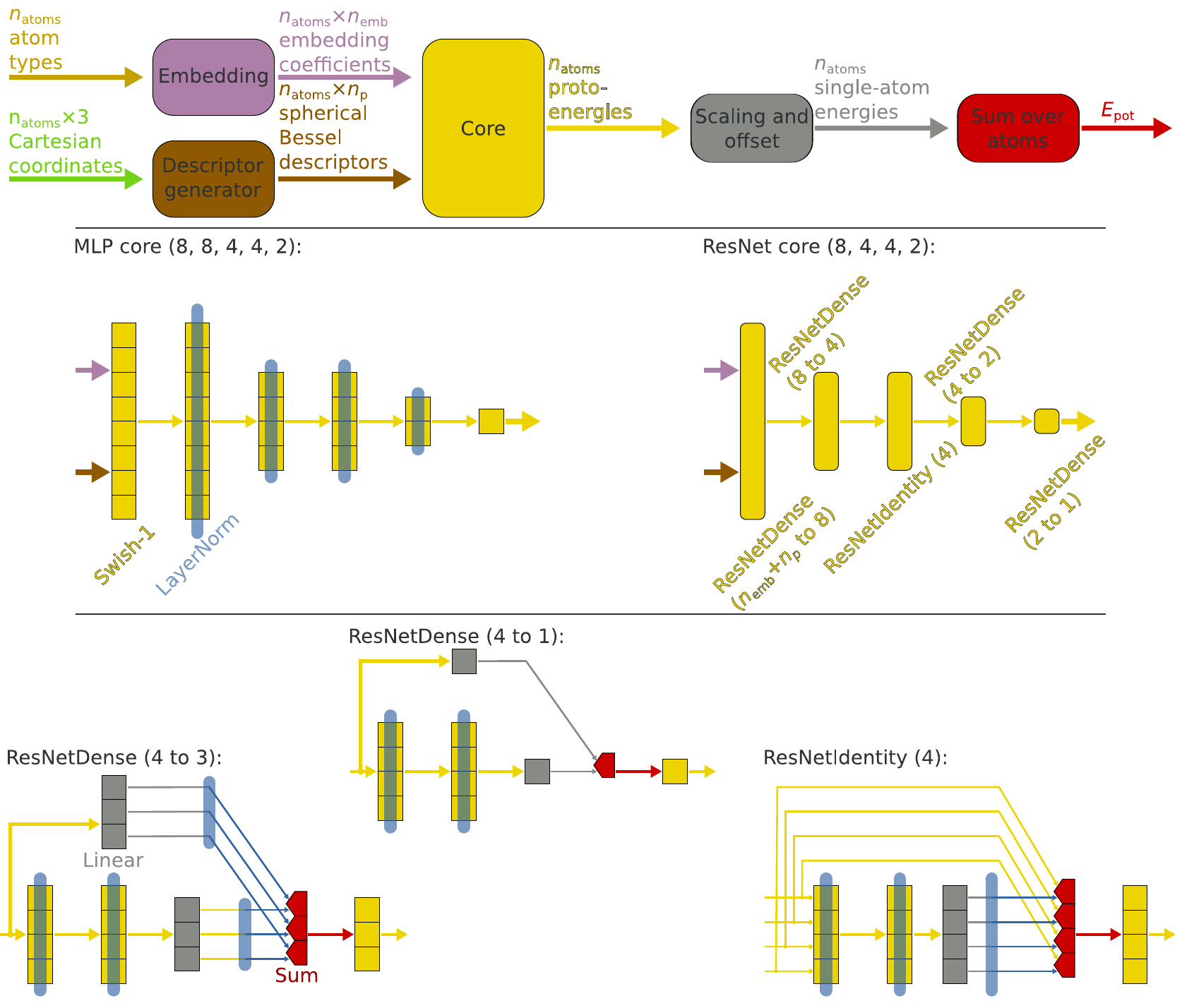}
    \caption{Top: schematic description of the NeuralIL architecture (basic homoscedastic case). Center: NNs making up the core of the model in the original NeuralIL (left) and the new ResNet-based version (right). Bottom: structure of the basic blocks of the ResNet for regression, namely the general dense block, the special case with output size 1, and the identity block. The numbers of neurons in these example diagrams are kept smaller than in the actual model to facilitate their interpretation. A single arrow between a pair of layers denotes all-to-all connections. A LayerNorm block superimposed on a layer of neurons is to be interpreted as acting on the linear combination of the inputs plus the offset, before it is passed into the activation function.}
    \label{fig:blocks}
\end{figure*}

The starting point for all of our models is NeuralIL, introduced in Ref.~\onlinecite{MontesCampos_JCIM22} and represented schematically in the top panel of Fig.~\ref{fig:blocks}. In terms of data flow and enforcement of the fundamental symmetries, NeuralIL falls in the class of NNFFs originally developed by Behler and Parrinello\cite{behler2007generalized} on heuristic grounds. However, the design and implementation takes advantage of many later advances from the mainstream ML world to improve performance and robustness.

A calculation starts with the encoding of the set of atomic coordinates into atom-centered descriptors that are invariant with respect to global rotations and translations. Specifically, the relative positions of the  neighbors within a finite cutoff radius of each atom in the system are turned into an array of second-generation spherical Bessel descriptors.\cite{kocer2020continuous} Besides the cutoff radius, the only parameter that needs to be chosen is the maximum radial order ($n_\mathrm{max}$) of the basis functions. The density is encoded independently for each pair of chemical elements, and for a system containing atoms from $n_{\mathrm{el}}$ of them, the total number of descriptors per atom is $n_p=\pqty*{n_\mathrm{max}+1}\pqty*{n_\mathrm{max}+2}\times n_{\mathrm{el}}\pqty*{n_\mathrm{el}+1}/4$. Not explicitly depicted in Fig.~\ref{fig:blocks} is the fact that periodic boundary conditions along all or some of the directions of space are taken into account in this phase, which requires up to $9$ more quantities to define the relevant box vectors. The spherical Bessel descriptors do not explicitly encode the chemical identity of the atom at the center of each environment, so we supplement them with $n_{\mathrm{emb}}$ embedding coefficients depending only on that atom's type.

The concatenated array of embedding coefficients and descriptors is fed into an NN, hereby called the \emph{core model}, with a single scalar output per atom that we call the \emph{proto-energy}. The proto-energies are then multiplied by a common trainable factor and displaced by a common trainable offset before being added together to form the final prediction of the potential energy of the atomic configuration. This sum over atoms is the last step in enforcing the physical symmetries of the model, since it removes the dependence on the arbitrary order of the atoms in the input. Although originally introduced heuristically, it constitutes a particular instance of a \emph{set pooling} or \emph{deep sets} architecture, whose representative power has been systematically studied later.\cite{deep_sets, deep_sets_limitations, bueno2021on} In this more modern light, the sum over atoms is not the only possible kind of pooling layer,\cite{equilibrium_aggregation} but it is still a physically appealing choice to achieve the right scaling of the total potential energy with the number of atoms. The interpretation of the scaled and displaced proto-energies as single-atom energies in a general physical context is dubious -- for instance, two rounds of training of the same model on the same data can lead to almost identical total energies while predicting only weakly correlated single-atom energies. However, having them add up to the total energy is enough to enable their use whenever an appropriate gauge principle holds, as is the case in MD thermal transport calculations.\cite{Ercole2016}

In the earliest implementations of this kind of NNFF, the core model was a fully connected MLP with traditional sigmoid activation functions. Despite the known theoretical representation power of those NNs, they are also famously prone to phenomena like vanishing gradients and dead neurons. Those result in slow training, drastic limitations to the practical depth of the NN, and a need for fine tuning to avoid divergences and overfitting. Moving beyond the constraints of the sigmoid-based MLP has been a significant contributor to the success of modern NN-based ML in general, and a lot of effort continues to be directed at improving all the basic building blocks of modern models. The original NeuralIL still used an MLP as its core model, but incorporated several of those key improvements: instead of a sigmoid-like activation function, it uses Swish-1, a state-of-the-art differentiable alternative,\cite{swish} and it makes aggressive use of normalization (specifically LayerNorm\cite{LayerNorm}) between layers whenever possible. A schematic depiction of that core model can be found in the center-left panel of Fig.~\ref{fig:blocks}; note that neither the inputs nor the final layer are normalized, the former so as not to destroy the scale of the atomic density and the latter because it contains a single neuron.

Unfortunately, even this improved MLP eventually hits its limit. For demanding data sets, the probability of a suboptimal training run can be significant enough that an average of one model in an ensemble of five or ten might be an outlier with anomalously large errors. Moreover, decreasing returns are obtained by adding extra layers to the basic MLP design.

To overcome this situation, we completely scrap the MLP and replace it with a variation on the idea of a ResNet. ResNets were designed specifically to enable revolutionarily deep convolutional architectures\cite{ResNet} of up to 1000 layers while avoiding pitfalls like vanishing gradients. The idea behind them is that, while in principle a deep NN could spontaneously \enquote{shed} some of its layers by training them to reproduce the identity function, its practical functional form makes such a state unlikely to be reached. On the other hand, it is very easy to train a layer to ignore part of its inputs by driving the required coefficients to zero. Therefore, instead of a single path through fully connected layers, ResNets are built out of units where information is sent along both a deep and a shallow path, whose results are combined in a trainable manner in order to generate the final output. The viability of ResNet-inspired connections for NNFFs has been illustrated before, using a convolutional architecture.\cite{schutt2017schnet} Here we base our approach to residual learning on the adaptation of ResNets to deep regression problems presented in Ref.~\onlinecite{RegressionResNet}, but we replace BatchNorm\cite{BatchNorm} with LayerNorm due to the complications that the former introduces in the calculation of derivatives. This regression ResNet is made up of two kinds of blocks: identity blocks ($N$ to $N$ mappings) and dense blocks ($N$ to $M\ne N$ mappings). The bottom panel of Fig.~\ref{fig:blocks} shows both, along with the slightly special case of a dense block with $M=1$, and its center-right panel shows an example of a ResNet core built out of those blocks.

We have found\cite{sebastian_LJ} that for certain high-energy configurations not typically found in training sets (e.g. those with some very small interatomic distances) adding a simple pair potential can help avoid unphysical behavior and stabilize MD simulations. Therefore, a Morse contribution of the form

\begin{equation}
E_{\mathrm{Morse}}\pqty*{r} = \sum\limits_{i,j\in\mathrm{atoms}} V_{\mathrm{Morse}}\pqty*{r_{ij}}\text{, with }V_{\mathrm{Morse}}\pqty*{r}=De^{-A\pqty*{r - B}}\sqty*{e^{-A\pqty*{r - B}} - 2}
\label{eqn:morse}
\end{equation}

\noindent can optionally be added to the NeuralIL potential energy. Here the $A$, $B$ and $D$ parameters for each pair of elements are derived, using Kong's mixing rules, from element-specific $A$, $B$ and $D$. In turn, each of those is calculated from a fully trainable real parameter that is passed through SoftPlus function to constrain the result to positive values. The Morse component does not impose any limit on the total potential energy function, since the $D$ parameters can still be driven to zero during training; moreover, we multiply each of the pair contributions by a completely smooth cutoff based on the bump function\cite{tu2010introduction} to avoid any discontinuity of the forces or their derivatives. The cutoff radius $r_{\mathrm{cut}}$ is the same as for the descriptor generator, and the switching radius is a trainable parameter between $0.5r_{\mathrm{cut}}$ and $r_{\mathrm{cut}}$.

The full list of trainable parameters therefore comprises the embedding coefficients for each atom type, the weights and biases of all neurons in the core, the scaling factor and offset for transforming the proto-energies into single-atom energies, the element-specific Morse parameters and the switching radius of the Morse potential. In this article we use that Morse contribution only in the case of \ce{SrTiO3}.

NeuralIL is implemented on top of JAX,\cite{jax2018github} a Python library providing just-in-time (JIT) compilation and advanced automatic differentiation features. The former affords high performance and parallelization on both CPUs and GPUs. Thanks to the latter, vector-Jacobian and Jacobian-vector product operators (VJP and JVP, respectively) can be created automatically for code implemented in JAX, from which very efficient and accurate differentiation workflows are easy to assemble. Specifically, all forces in a given configuration are calculated through a single invocation of the VJP of the potential energy.\cite{MontesCampos_JCIM22} Physically interesting higher-order derivatives, such as the Hessian required for a normal mode calculation, can also be obtained without the even more significant coding overhead they would normally require. Likewise, the gradients used in training and in the active learning procedures described below are extracted through automatic differentiation. In addition to JAX itself, we use FLAX\cite{flax} to simplify model construction and parameter bookkeeping. Our code is distributed as open source and our models are available for download.\footnote{Zenodo record with DOI: \href{https://doi.org/10.5281/zenodo.7643625}{10.5281/zenodo.7643625}}

\subsection{Extension of the model to account for heteroscedasticity}

Building deep ensembles demands a heteroscedastic formulation of the regression problem and therefore a variance estimate. The extreme representation power of NNs offers the interesting opportunity to have the model itself produce this estimate and to train both components simultaneously. Although in principle this could be accomplished with a completely separate NN with common inputs, there are advantages to a higher level of integration. First, a scalar variance must satisfy the same symmetries as the main output of the FF (the potential energy) with respect to transformations of the system of coordinates and to permutations of atoms; second, it is reasonable to assume that the main outputs of the NNFF (potential energy and forces) and the variance are affected by common influence factors best expressed by quantities derived from the descriptors but with an intermediate level of elaboration. Taking this into account, we adopt the common strategy of notionally splitting the ResNet-based core model into a \emph{feature extractor} containing the layers closer to the input and a \emph{head} containing those closer to the output. We then add a second head to have the core generate an additional output, which is later scaled and displaced by trainable constants, filtered through a SoftPlus function to guarantee non-negativity, and summed over atoms to provide an estimate of $\sigma_E^2$, the variance of the total potential energy.

\begin{figure*}
    \centering
    \includegraphics[width=\textwidth]{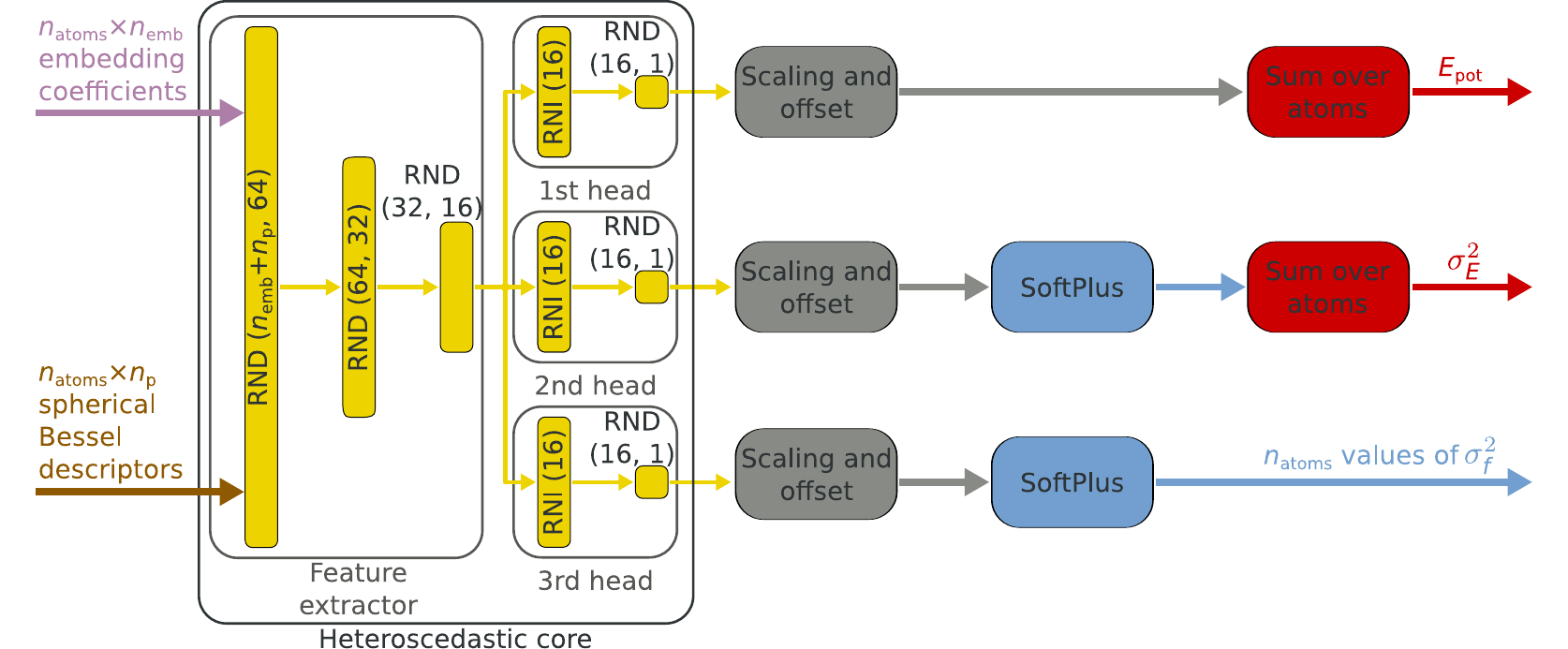}
    \caption{Schematic description of our extension of the NeuralIL architecture to allow for a heteroscedastic loss. Abbreviations: RNI = ResNetIdentity; RND = ResNetDense. The dimensions of the layers correspond to the actual model in this paper.}
    \label{fig:heteroscedastic}
\end{figure*}

Building $\sigma_E^2$ as a sum over atoms takes advantage of the set-pooling architecture to enforce the  permutation invariance, but the individual contributions to this quantity should not be interpreted as variances of the single-atom energies, nor should this construction be read as an assumption of a lack of correlation between the single-atom contributions to the energy. A better starting point for an atom-resolved heteroscedastic formulation is to attach a variance to each of the atomic forces. Despite the fact that all components of the force vector are obtained from the total energy by automatic differentiation, it is cumbersome to compute the variance of the force on atom $i$ along Cartesian axis $\alpha$, $\sigma^2_{f_i\supar{\alpha}}$, from derivatives of statistics of the energies -- as shown in Appendix~\ref{apx:covariance}, this requires a two-point calculation to estimate the correlation between the energies of two configurations. Instead, we add a third head to the ResNet core whose output is also passed to a scaling and offset neuron and a SoftPlus function, but not summed over atoms, and take that single atom result as an isotropic $\sigma^2_f$. A schematic representation of the heteroscedastic NeuralIL core with its three heads is presented in Fig.~\ref{fig:heteroscedastic}.

In contrast with the aforementioned situation regarding $\sigma_E^2$, the fact that the $\sigma^2_f$ thus calculated is a function of the descriptors centered at a single atom does place some constraints on the result (e.g. it is short ranged). Nevertheless, even in conjunction with the isotropic ansatz this leaves plenty of flexibility to obtain a good enough set of variances to derive point weights for a heteroscedastic model.

The core of our homoscedastic models use a ResNet architecture with composite layers of widths $64$, $32$ and $16$. The heteroscedastic extension uses the same sequence widths for the feature extractor and adds a ResNetIdentity($16$) layer for each of the three heads, as represented in Fig.~\ref{fig:heteroscedastic}

\subsection{Loss and training}\label{subsec:training}

After random initialization of all coefficients from a truncated normal distribution following the LeCun normal initialization,\cite{LeCuBottOrrMull9812} our homoscedastic models are trained by minimizing the following loss function:

\begin{equation}
    \begin{aligned}
    \mathcal{L} =& \frac{1}{2}\aqty*{\frac{0.2}{n_{\mathrm{atoms}}}\sum\limits_{i=1}^{n_{\mathrm{atoms}}} \log\sqty*{\cosh\pqty*{\frac{\norm*{\bm{f}_{i,\mathrm{predicted}} - \bm{f}_{i,\mathrm{reference}}}_2}{\SI{0.2}{\electronvolt\per\angstrom}}}}} \\
    +&\frac{1}{2}\aqty*{0.02 \log\sqty*{\cosh\pqty*{\frac{E_{\mathrm{pot}} - E_{\mathrm{pot},\mathrm{reference}}}{n_{\mathrm{atoms}}\times\SI{0.02}{\electronvolt\per\atom}}}}}.
    \end{aligned}
    \label{eqn:homoscedastic_loss}
\end{equation}

\noindent Here, $\aqty*{\cdot}$ denotes an average over configurations, estimated using a different minibatch at each iteration within an epoch. The log-cosh function smoothly interpolates between a quadratic and a linear regime, reducing the weight of outliers and providing a form of gradient clipping during training. Although the factors above have been arranged in such a way as to make the loss dimensionless, the formula dimensionally homogeneous, and its limits easy to identify, we have found that the relative weights of the energies and forces can be varied within a wide range with negligible effect on the results, and thus do not need to be tuned, when the training procedure laid out in this section is followed. Some of us have shown \cite{MontesCampos_JCIM22} that the forces alone contain enough information to drive the training of an NNFF, with the caveat that the origin of energies need to be adjusted at the end of the process. The role of the energy contribution to the loss in Eq.~\eqref{eqn:homoscedastic_loss} is mainly to render that additional step unnecessary.

On the other hand, our heteroscedastic loss has a hybrid structure:

\begin{equation}
\begin{aligned}
    \mathcal{L} &= \frac{1}{2}\aqty*{\frac{1}{2} \frac{\pqty*{E_{\mathrm{pot}} - E_{\mathrm{pot},\mathrm{reference}}}^2}{\sigma_{E}^2} + \log \pqty*{\frac{\sigma_{E}^2}{\SI{1}{\electronvolt^2}}}}\\
    & + \frac{1}{2}\aqty*{\frac{1}{n_{\mathrm{atoms}}}\sum\limits_{i=1}^{n_{\mathrm{atoms}}}\log\sqty*{\cosh\pqty*{\frac{\pi}{2}\frac{\norm*{\bm{f}_{i,\mathrm{predicted}} - \bm{f}_{i,\mathrm{reference}}}_2}{\sigma_{f,i}}}}+\log\pqty*{\frac{\sigma_{f,i}}{\SI{1}{\electronvolt\per\angstrom}}}}.
\end{aligned}
\label{eqn:heteroscedastic loss}
\end{equation}

\noindent The energy contribution has the well known form of the negative logarithm of a likelihood of a Gaussian distribution, as would be used in a conventional stochastic maximum-likelihood estimation. The force contribution can be interpreted as a gradient-clipped version of the same construction or as derived from the negative log-likelihood of a hyperbolic secant distribution\cite{Fischer2013-wh} (a super-Gaussian distribution with an excess kurtosis of $2$). The reason for the different treatment of the energy is that, even if we were to assume a similarly leptokurtic distribution for each of the atomic energies, the total energy would still be very close to a Gaussian.

The standard method for minimizing the loss in NN training is stochastic gradient descent (SGD). In the basic incarnation of this method, each parameter of the model is updated by an amount proportional to the corresponding component of the local gradient of the loss computed using a minibatch. Refined versions of SGD, like the popular ADAM\cite{ADAM} and derived methods, keep a limited memory of moments of the gradient to stabilize and accelerate the descent toward the minimum. Here we use VeLO, a fully nonlinear optimizer that takes a radical departure from the philosophy of SGD to achieve much higher performance.\cite{VeLO} VeLO is itself a complex ML model that computes the update to each parameter at each iteration based on the current values of all parameters and the gradient of the loss. That model has been meta-trained, using evolutionary algorithms, on a wide sample of models including MLPs, ResNets, convolutional networks, transformers and autoencoders. We use the stochastic optimization library OPTAX,\cite{deepmind2020jax} integrated both in the JAX ecosystem and with the open-source implementation of VeLO. A remarkable feature of VeLO is that it renders the choice of a learning rate unnecessary: its only parameter is the total number of epochs.

\subsection{Bootstrap aggregation, committees and deep ensembles}

To build a bootstrap-aggregation ensemble, we repeat the following process $10$ times: we sample, with replacement, a pool of a size equal to $75\%$ of the total data set; we then take $50\%$ of that pool as our training data and  leave the remaining $50\%$ for validation; finally, we train a single NNFF based on those subsets. In terms of performance this comes close to being equivalent to training the basic NNFF ten times,\cite{Zhu_arXiv23} with the only savings coming from the smaller training and validation data sets and from the avoided JIT recompilations. This stands in total contrast with the situation during inference, where evaluating the ten NNs for each configuration imposes very little overhead in comparison with evaluating only one because the most time-consuming part of the calculation, getting the descriptors and the associated VJP, is only performed once.

Committees take advantage of this imbalance of computational complexity by training all networks on the same data, the only source of randomness being the different initialization of the coefficients of the NN. We implement a committee as a single FLAX model with an additional dimension for each tensor, which runs over the members of the ensemble. Accordingly, each $10$-member committee used in this work is efficiently trained in a single run with little performance loss compared to the training of a single NN. Note that this optimization does introduce a subtle reduction in diversity because the splitting of the training data in minibatches will also be the same for all members of the committee. The assumption of homoscedasticity results in the well-known average and uncertainty estimates for bootstrap-aggregation ensembles and commitees
\begin{subequations}\label{grp:unweighted}
\begin{gather}
    \aqty*{E_{\mathrm{pot}}}_{\mathrm{HOS}} = N^{-1}_{\mathrm{ensemble}} \sum\limits_{i\in\mathrm{members}} E_{\mathrm{pot},i}\label{eqn:averages:unweighted}\\
        \sigma^2_{E,\mathrm{HOS}} = \frac{1}{N_{\mathrm{ensemble}}\pqty*{N_{\mathrm{ensemble}}-1}} \sum\limits_{i\in\mathrm{members}} \pqty*{E_{\mathrm{pot},i}-\aqty*{E_{\mathrm{pot}}}_{\mathrm{HOS}}}^2.\label{eqn:variances:unweighted}
\end{gather}
\end{subequations}

The deep ensembles share the computational advantage with the committees that they can be implemented as a single model. The increase in complexity as a consequence of the additional heads is barely noticeable in terms of computational time, which is always dominated by the descriptor calculation. However, assuming a lack of correlation between members, the additional head (see Fig.~\ref{fig:heteroscedastic}) allows us to express a collective prediction from a deep ensemble using the minimum-variance linear unbiased estimator of the corresponding population mean, the weighted average 
\begin{subequations}\label{grp:weighted}
\begin{align}
    \aqty*{E_{\mathrm{pot}}}_{\mathrm{HES}} &= \pqty*{\sum\limits_{i\in\mathrm{members}} \frac{1}{\sigma^2_{E,i}}}^{-1} \sum\limits_{i\in\mathrm{members}} \frac{E_{\mathrm{pot},i}}{\sigma^2_{E,i}} \label{eqn:averages:weighted}\\
    \sigma^2_{E,\mathrm{HES}} &= \pqty*{\sum\limits_{i\in\mathrm{members}} \frac{1}{\sigma^2_{E,i}}}^{-1}\label{eqn:variances:weighted}\end{align}
\end{subequations}

\noindent For deep ensembles, however, the standard prescription for uncertainty estimation\cite{lakshminarayanan2017simple,AutoDEUQ} combines an unbiased committee variance as a proxy for epistemic uncertainty with the average of the additional head as an approximation to aleatoric uncertainty:

\begin{equation}
    \sigma^2_{E,\mathrm{DE}} = \underbrace{\vphantom{\sum\limits_{i\in\mathrm{members}}}N_{\mathrm{ensemble}}\sigma^2_{E,\mathrm{HOS}}}_{\text{epistemic}} + \underbrace{N^{-1}_{\mathrm{ensemble}}\sum\limits_{i\in\mathrm{members}}\sigma^2_{E,i}}_{\text{aleatoric}}
    \label{eqn:variance:DE}
\end{equation}

\noindent 
In our heteroscedastic tests we use both $\sigma^2_{\mathrm{DE}}$ and $\sigma^2_{\mathrm{HES}}$ to see if there is an advantage to either of them. When employing $\sigma^2_{\mathrm{DE}}$, our estimate of the mean is  Eq.~\eqref{eqn:averages:unweighted} for consistency. By definition, $\sigma^2_{\mathrm{DE}}$ is larger than $\sigma^2_{\mathrm{HES}}$ or $\sigma^2_{\mathrm{HOS}}$ by a factor of approximately $N_{\mathrm{ensemble}}$, so we scale $\sigma_{\mathrm{DE}}$ by $\sqrt{N_{\mathrm{ensemble}}}$ for direct comparisons. We furthermore note that although $\sigma^2_{\mathrm{HES}}$ and the second term in $\sigma^2_{\mathrm{DE}}$ are termed aleatoric in literature, they predict uncertainty both based on error in the input data, and on the model itself if some datapoints are more difficult to fit than others. Since the input data in our study contains no error by construction, all contributions to the uncertainty are, strictly speaking, epistemic, i.e. caused by inadequacies of the model that can be reduced by adding more data.
We use analogous constructions for the uncertainties of the forces in the homo- and heteroscedastic cases.

\subsection{Data generation for EAN.}

The data used for training the EAN potential is the same set employed in the original NeuralIL article,\cite{MontesCampos_JCIM22} distributed as part of its supplementary information, and we choose the same training/validation split to make our new results more directly comparable with those from the original design. The data set comprises $368$ structures directly extracted from a classical MD trajectory plus $373$ others to which a DFT energy minimizer has been applied for five steps. All energies and forces had been calculated using the GPAW\cite{gpaw1,gpaw2} DFT package with a linear combination of atomic orbitals (LCAO) basis set; see the original reference for details of the parameterization.

We then run an MD simulation using a single NNFF trained on that data and the JAX-MD library\cite{jaxmd2020} to obtain $12$ MD trajectories in the $NVT$ ensemble at $T=\SI{298}{\kelvin}$ with a duration of \SI{80}{\pico\second} each and a time step of \SI{1}{\femto\second}. The thermostat used is a Nosé-Hoover chain\cite{Tuckerman_2006} with a length of $5$ and a time constant of $100$ time steps. The initial conditions for all $12$ trajectories are based on the same snapshot from the training set, but the velocities are initialized at random from a Maxwell-Boltzmann distribution at $T=\SI{298}{\kelvin}$ and are different in each case. We select $4$ out of $12$ trajectories as representative of different situations vis-à-vis uncertainty, as detailed in the next section. We delimit the high-uncertainty segments of those four trajectories and we sample $125$ equispaced points that we then run through the same DFT setup described above to obtain energies and forces to supplement the data set in a new round of training.

\subsection{Data generation for the \ce{SrTiO3} surface.}

To generate the reference data for bulk \ce{SrTiO3} we start from a $3\times 3\times 3$, $135$-atom cubic supercell. We employ the same lattice parameter as in Ref.~\onlinecite{STO_surfaces}, namely \SI{4.01}{\angstrom}. We then rattle the atomic positions using random displacements generated from an isotropic Gaussian distribution. In particular, the mass-weighted displacement $m_i u_i^{\pqty{\alpha}}$ of atom $i$ from its equilibrium position along Cartesian axis $\alpha$ is drawn from a Gaussian distribution with variance equal to $3\frac{T}{k_\mathrm{B}\theta_\mathrm{D}}$, where $k_{\mathrm{B}}$ is the Boltzmann constant and $\theta_\mathrm{D}$ is the Debye temperature of the material.  We take $\theta_\mathrm{D}$ as \SI{418.5}{\kelvin}, the average of the experimental values compiled in Ref.~\onlinecite{PhysRevMaterials.3.022001}, and we generate $600$ configurations with $T=\SI{500}{\kelvin}$ and a further $600$ with $T=\SI{1000}{\kelvin}$. We set aside $100$ points from each of those series to build a test set, and use the remaining $500$ for training and validation. We again employ GPAW in the LCAO mode, with the PBE approximation\cite{Perdew1996} to the exchange and correlation terms and $\Gamma$-only sampling in reciprocal space.

To progressively improve the performance of the \ce{SrTiO3} NNFF trained on these bulk data on the  $4\times 1$ \ce{SrTiO3}$(110)$ surface reconstruction we apply an active-learning strategy inspired by the work of Bombarelli and coworkers.\cite{Bombarelli_adversarial} Starting from a predefined configuration we first apply a random displacement drawn from a Gaussian distribution with zero mean and a standard deviation of \SI{0.1}{\angstrom} to each atom. We then run a numerical maximization of the logarithm of the adversarial loss
\begin{equation}
\mathcal{L}_{\mathrm{adv}} = \sigma^2_{\bm{f}}\exp\bqty*{-\frac{E_{\mathrm{pot}}}{k_{\mathrm{B}}T}}
\label{eqn:adversarial}
\end{equation}
for $T=\SI{500}{\kelvin}$ using the Powell conjugate-directions method,\cite{Powell} setting the maximum relative norm of the change in the independent variables acceptable for convergence to $x_{\mathrm{tol}}=\num{100}$. This allows us to explore regions of configuration space that are both unknown and physically plausible. We initially generate $100$ surface configurations using this method starting from the same initial positions used to start the evolutionary search presented Ref.~\onlinecite{STO_surfaces}. We then run a second iteration where each new point is generated starting from a randomly chosen configuration from the first iteration.

As test data for the \ce{SrTiO3} surface we use $5000$ structures chosen at random from a larger collection of $29500$ gathered from three different $500$-generation stochastic trajectories of the covariance matrix adaptation evolution strategy (CMA-ES) search algorithm.\cite{STO_surfaces} Those three trajectories differ in the number of layers of atoms that are allowed to move during the optimization. The atomistic model used to study these reconstructions contains $136$ atoms,  employs periodic boundary conditions only in the directions perpendicular to $(110)$, and is always kept symmetric with respect to its central plane. It therefore simulates a slab with two equivalent free surfaces.

\section{Results and discussion}

\subsection{Performance of the revised NeuralIL on EAN data}

We first examine the general performance of the models in terms of speed and accuracy using the EAN data set. As a reference, the first row of Table~\ref{tbl:EAN_comparison} contains the mean average errors in the prediction of the total potential energy and the forces on the validation portion of that data set using the original NeuralIL model from Ref.~\onlinecite{MontesCampos_JCIM22}. The next block of three rows lists the values of the same statistics for the main type of model employed in this manuscript (with a ResNet core and the VeLO learned optimizer) with a different number of total training epochs and therefore a different CPU budget for the optimization phase. All three showcase the enormous performance advantage of the new architecture and procedure: even with only $21$ epochs, the MAE values are competitive, and with $101$ epochs the final model outperforms the original for both energy and force predictions. We settle on $51$ training epochs as an excellent compromise. This amounts to a $10$- to $20$-fold acceleration with respect to the original $500$ to $1000$ training epochs.

\begin{figure}
    \centering
    \includegraphics[width=.6\columnwidth]{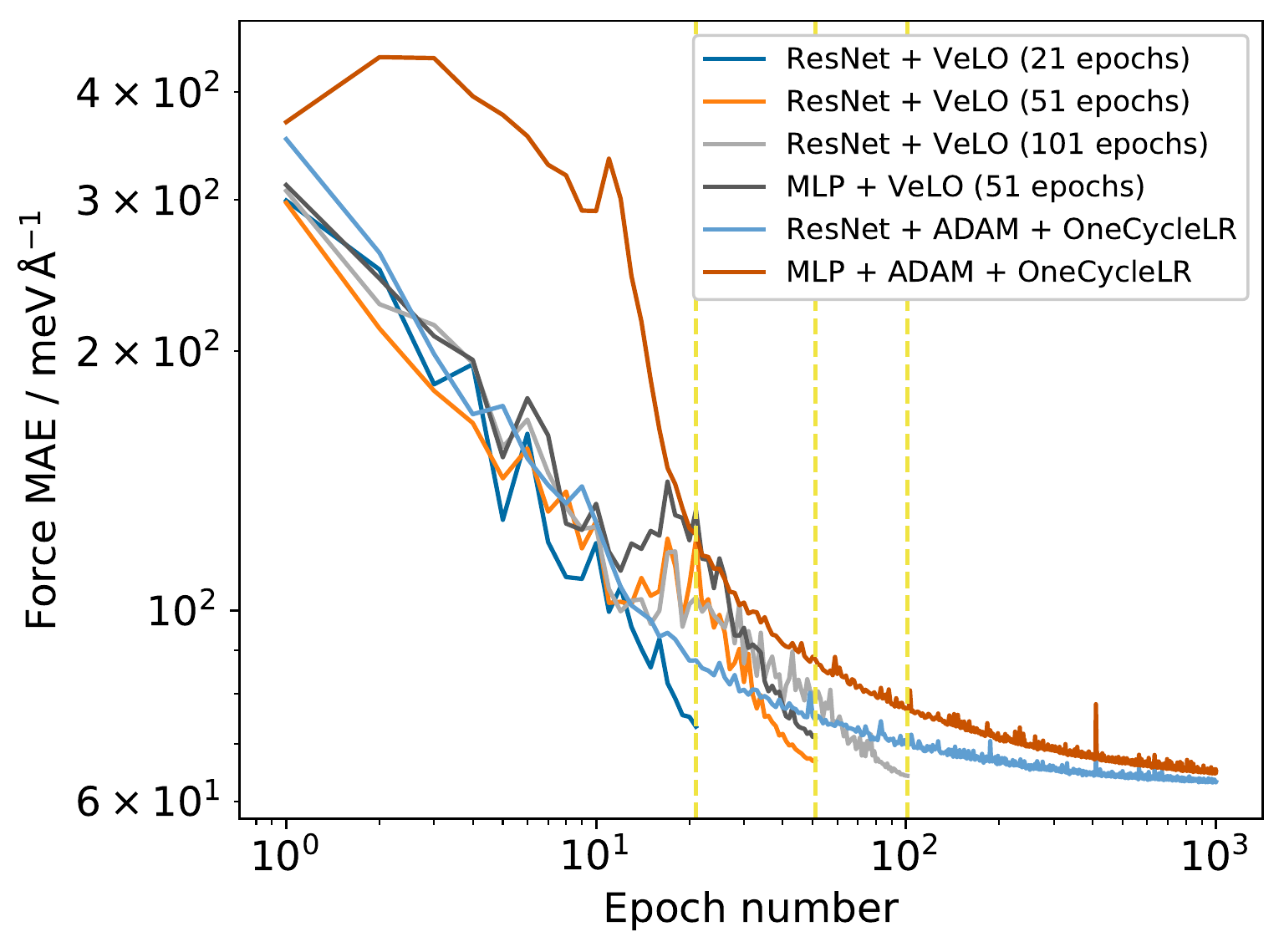}
    \caption{Evolution of the validation loss for the EAN data set with different choices of core architecture and optimizers. The vertical lines are placed at the $21$-, $51$- and $101$-epoch marks.}
    \label{fig:resnet_and_velo}
\end{figure}

To help identify which elements are determinant for this improvement, we also analyze the results of two hybrid models: one keeps the MLP core of NeuralIL but uses the VeLO optimizer for training while the other employs the new ResNet core along with the original Adam optimizer with a one-cycle learning-rate schedule. Naturally, both of them show competitive MAEs (see the third block of rows in Table~\ref{tbl:EAN_comparison}). A a more detailed picture emerges from the evolution of the loss during training as depicted in Fig.~\ref{fig:resnet_and_velo}. Looking at the value of the loss at $51$ epochs it can be that either of the major upgrades to the original design is, in isolation, enough to achieve a dramatic improvement in performance after a short training period, and could potentially enable an active learning workflow by itself. Our experience with more challenging data sets points to VeLO often having an edge. However, the combination of both techniques always offers an even better result with marginal additional cost.

\begin{table}
\begin{tabular}{lcccc}
Model & MAE $E_{\mathrm{pot}}$ & MAE $\bm{f}$ \\
 & (\si{\milli\electronvolt\per\atom}) & (\si{\milli\electronvolt\per\angstrom}) \\\hline
NeuralIL & $1.86$ & $65.6$ \\\hline
ResNet + VeLO (21 epochs) & $1.37$ & $73.2$ \\
ResNet + VeLO (51 epochs) & $1.21$ & $66.8$ \\
ResNet + VeLO (101 epochs) & $1.47$ & $64.2$ \\\hline
ResNet + ADAM + OneCycleLR & $1.35$ & $63.5$ \\
(1001 epochs) & & \\\hline
Committee & & \\
(10 NNs, ResNet, VeLO, 51 epochs) & $1.16$ & $63.3$ \\
Deep ensemble & & \\
(10 NNs, ResNet, VeLO, 51 epochs) & $1.75$ & $65.0$ \\\hline
\end{tabular}
\caption{Validation statistics of the original NeuralIL force field,\cite{MontesCampos_JCIM22} the single models shown in Fig.~\ref{fig:resnet_and_velo}, a 10-NN committee and a 10-NN deep ensemble, all with the same EAN training and validation data sets.}
\label{tbl:EAN_comparison}
\end{table}

We also check that we do not incur a significant degradation in predictive power by using a committee or a deep ensemble instead of a single neural network. As the last two rows in Table~\ref{tbl:EAN_comparison} show, both afford comparable or better accuracy than a single NN with the same parameters.

\subsection{Assessment of uncertainty in EAN using committees and deep ensembles}

\begin{figure*}
    \centering
    \includegraphics[width=\textwidth]{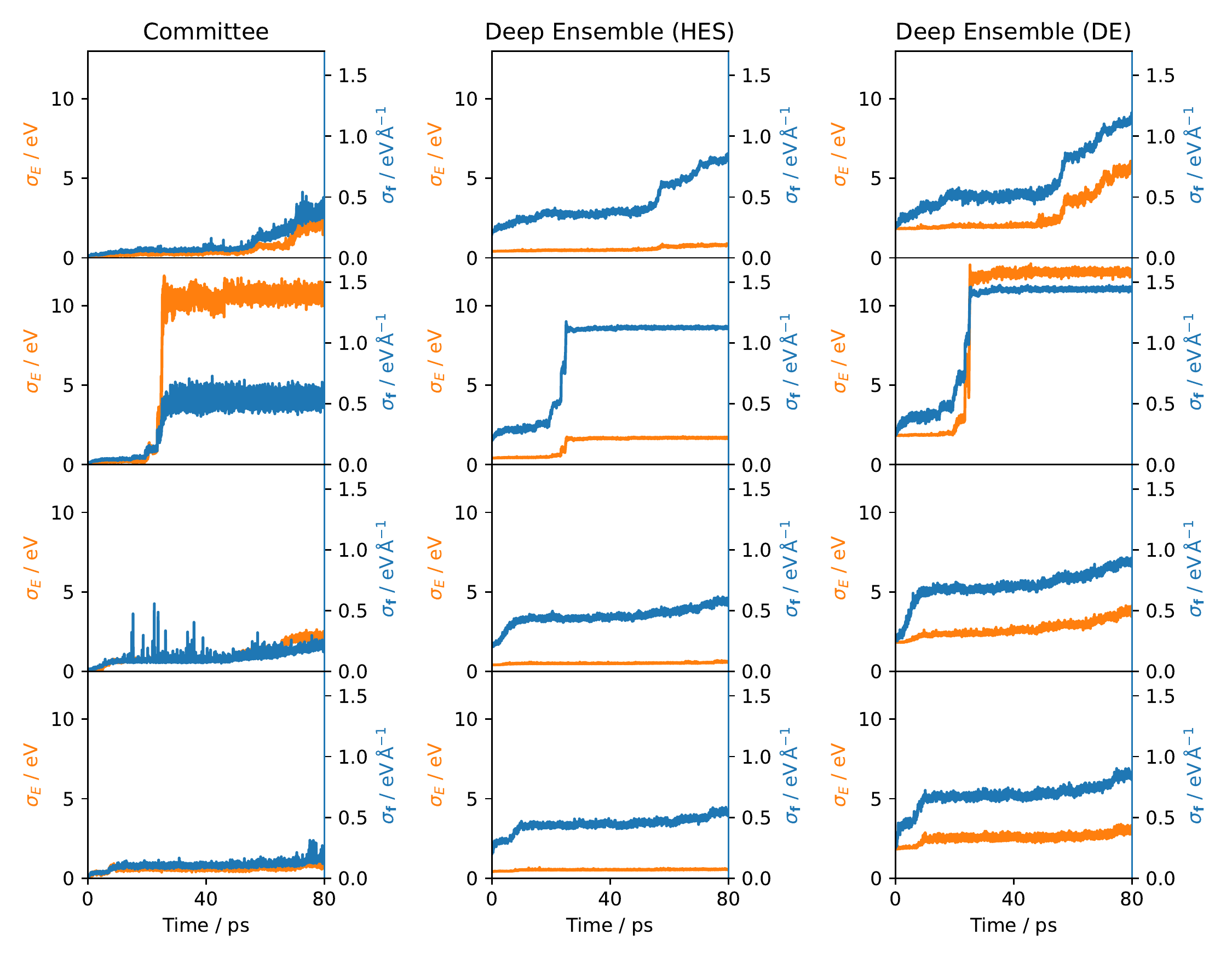}
    \caption{Evolution of the committee and deep-ensemble uncertainty metrics along a selection of four MD trajectories with different initial conditions, using a NNFF based on the training data set from Ref.~\onlinecite{MontesCampos_JCIM22}. The $\sigma_f$ for the whole system is defined as $\sqrt{\sum\limits_{\mathrm{atoms}}\sigma_f^2}$.}.
    \label{fig:ean_traj_analysis_a}
\end{figure*}

\begin{figure*}
    \centering
    \includegraphics[width=\textwidth]{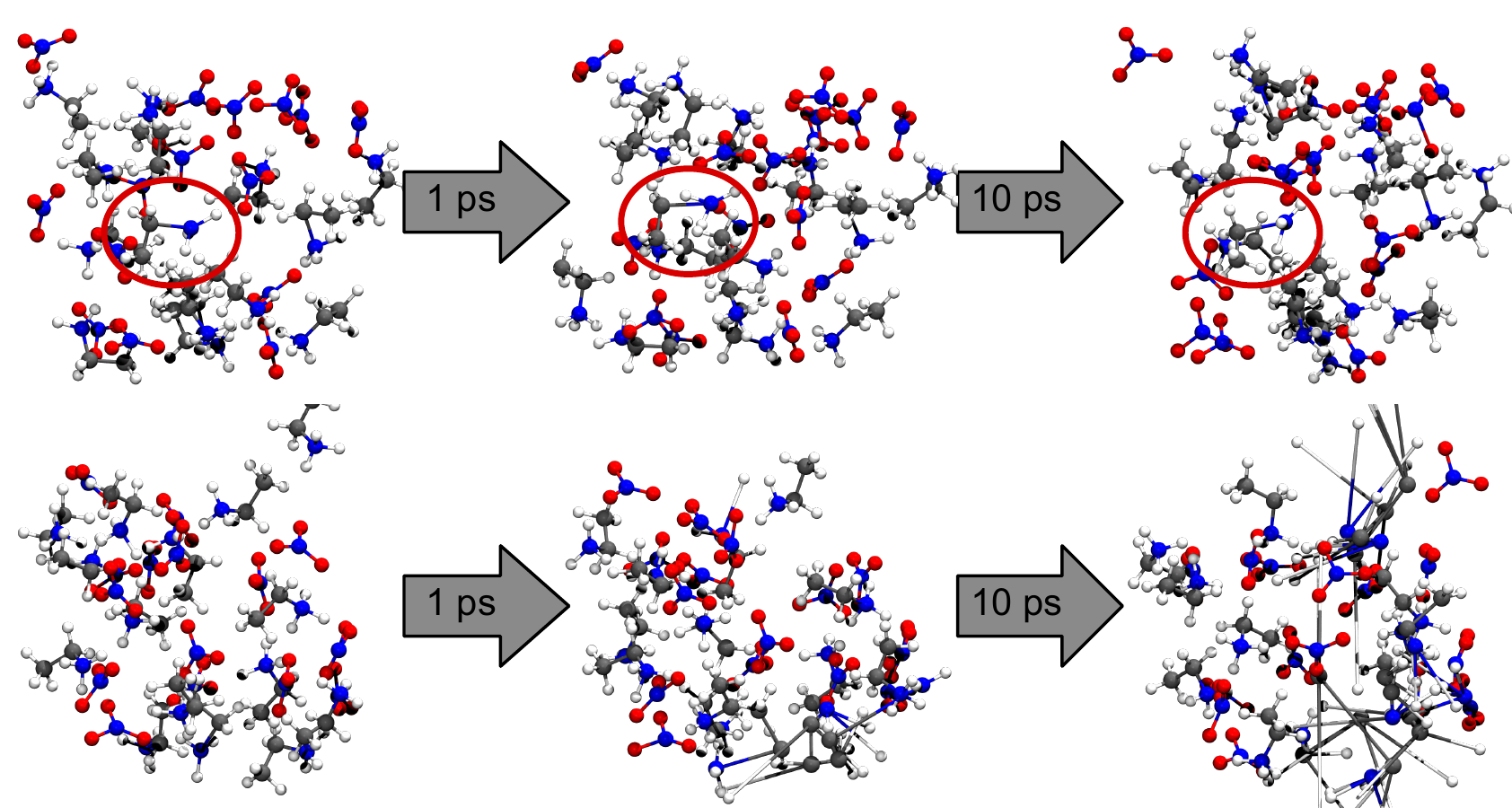}
    \caption{Sequences of snapshots illustrating the emergence of problems in the first two MD trajectories represented in Fig.~\ref{fig:ean_traj_analysis_a} when using a potential trained only on the original EAN data set. Gray, red, blue and white spheres represent carbon, oxygen, nitrogen and hydrogen atoms, respectively.}
    \label{fig:snapshots}
\end{figure*}

We now start our assessment of committees and deep ensembles as tools for uncertainty estimation by analyzing their behavior along a MD trajectory. Not only does it involve physically relevant collective changes of all degrees of freedom, but it also one of the prime targets for quality assessment and improvements of MLFFs. The EAN data set from Ref.~\onlinecite{MontesCampos_JCIM22} in particular makes for an interesting test case since the configurations therein were extracted from classical molecular mechanics trajectories but the energies and forces were calculated with DFT; therefore, an MD simulation run with the trained potential will not retrace the steps of the original classical FF but explore other regions. As could be expected, this is immediately visible in the uncertainty derived from a committee or a deep ensemble. Figure~\ref{fig:ean_traj_analysis_a} illustrates this point taking four example trajectories with very different characteristics. The two topmost ones can be described as pathological: in the first trajectory, an anomalous configuration of the H atoms in a cation causes bond breakage in a local but progressively growing environment; in the second, the interaction between two cations leads to an artifactual dissociation of a \ce{C-N} bond in each of them that causes a catastrophic chain effect where other ions also lose their structural integrity. Representative sequences of snapshots for these two cases are presented in Fig.~\ref{fig:snapshots}. The other two trajectories are comparatively uneventful and merely involve the transfer of a single hydrogen from a cation to an anion; this process is physically plausible in a protic ionic liquid but is not represented in the training data. Although the global trend of the uncertainty with time is monotonically increasing in all cases, showing the aforementioned evolution towards unexplored configurations, the more catastrophic events leave a clear fingerprint in the plots and enable an easy classification of the trajectories. There is no clear advantage to the more advanced deep ensembles with respect to the simpler committees. Just like the forces contain more information than the total energy, the uncertainty in the forces is more informative and better behaved than its counterpart for the total energy.

\subsection{Improving the potential to obtain more accurate MD trajectories}

\begin{figure*}
    \centering
    \includegraphics[width=\textwidth]{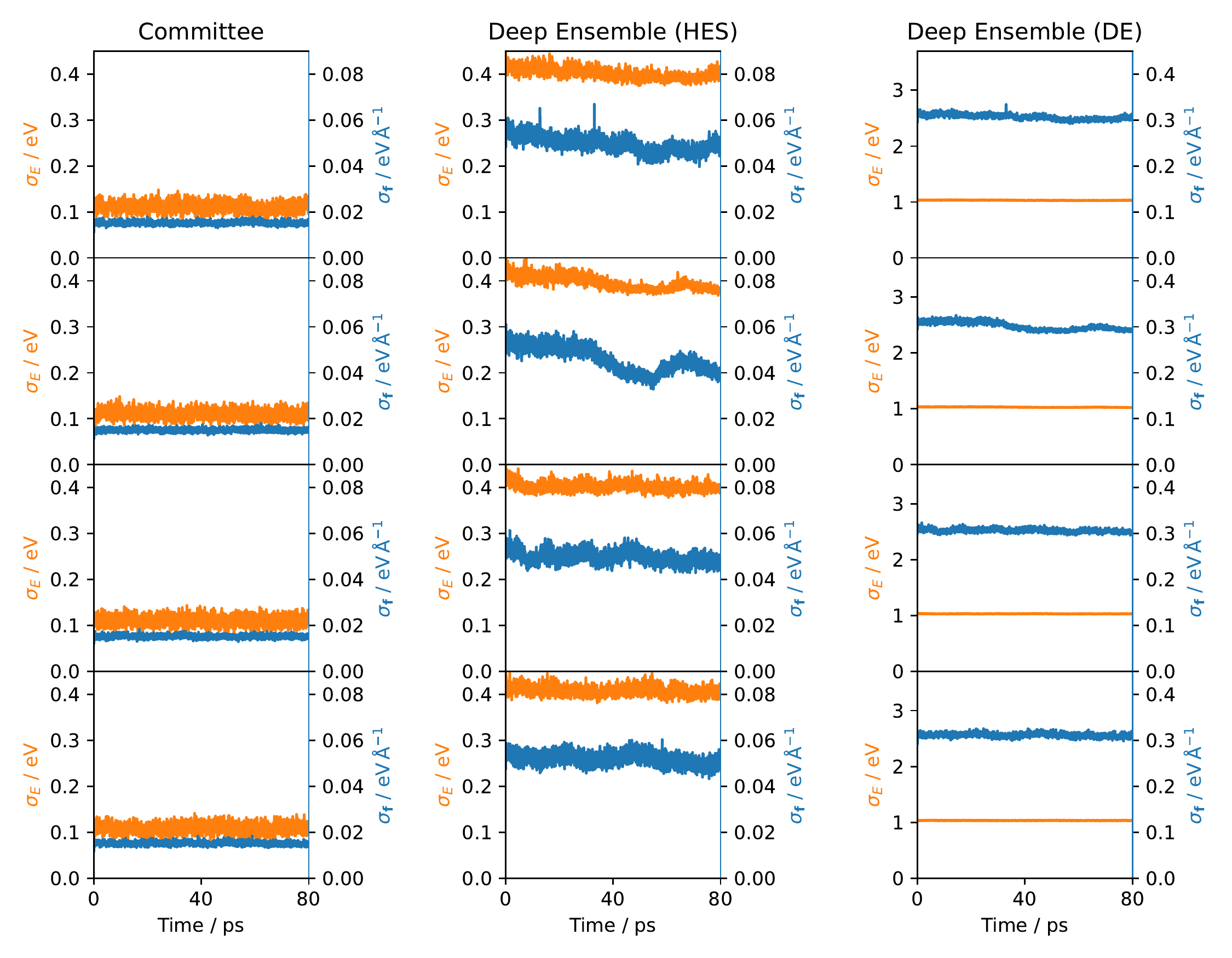}
    \caption{The same metrics as Fig \ref{fig:ean_traj_analysis_a}, evaluated on MD trajectories started from the same initial conditions but driven by an FF whose training data set was supplemented with points from the high-uncertainty segments of the trajectories in Fig \ref{fig:ean_traj_analysis_a}}.
    \label{fig:ean_traj_analysis_b}
\end{figure*}

\begin{figure}
    \centering
    \includegraphics[width=0.6\columnwidth]{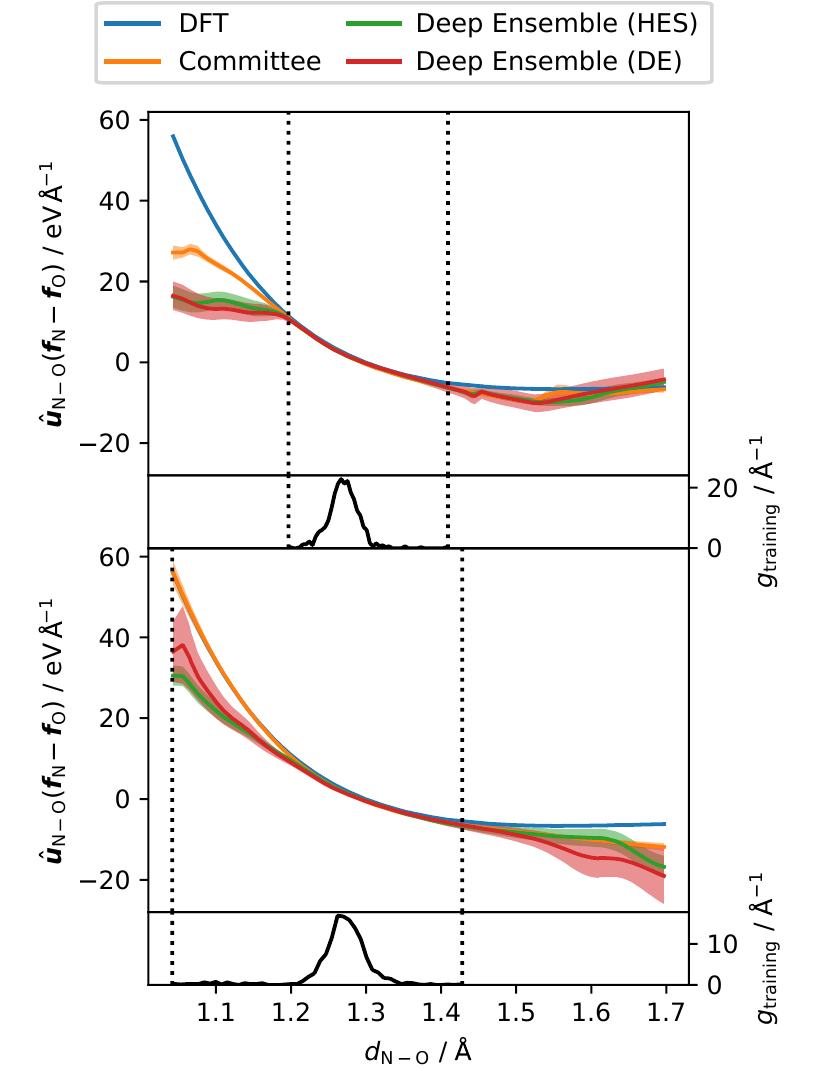}
    \caption{Committee and deep-ensemble predictions of the force along an N-O bond with the associated uncertainties  using the original EAN data set (top) and the version supplemented with high-uncertainty MD data (bottom). Uncertainties are represented as a filled area with a half-width equal to the standard deviation $\sigma_{\mathrm{HOS}}$, $\sigma_{\mathrm{HES}}$, and $\sigma_{\mathrm{DE}}/\sqrt{10}$. Also depicted are the DFT ground truth and a frequency density plot of the corresponding distance among the training data, with the minimum and maximum of the latter indicated as vertical dotted lines.}
    \label{fig:bond_stretching}
\end{figure}

The fact that our MD trajectories stray into inadequately explored segments of configuration space is detectable in the uncertainties and causes spurious behavior that either makes the simulation crash or renders its results unphysical. To solve this, we retrain the NNFF after supplementing the data set with the $500$ snapshots extracted from the high-uncertainty segments of the trajectories as explained in the methods section. We then restart the four example MD trajectories from the same initial conditions and monitor the uncertainty estimates again. As desired, this has the effect of avoiding any unphysical events along the trajectories and, as shown in Fig.~\ref{fig:ean_traj_analysis_b}, this is reflected in a stabilization of all uncertainty metrics at values close to their starting point. The change is also obvious in the analysis of an individual degree of freedom with the retrained potential (Fig.~\ref{fig:bond_stretching}, bottom panel): the retrained deep ensemble, but especially the retrained committee approximate the DFT results in a significantly wider range.

The uncertainty estimates are also analyzed for a series of configurations sampled along a trajectory determined by the change in a single degree of freedom in Figure~\ref{fig:bond_stretching}. In particular, we study the stretching of a single N-O bond in an anion to establish a direct comparison with the bootstrap-aggregation strategy used in Ref.~\onlinecite{MontesCampos_JCIM22}. Figure~\ref{fig:bond_stretching}(top) shows that the ensemble-based approaches can provide a reliable proxy for error, which stays at manageable levels even at a certain distance from the values of those degrees of freedom that are better represented in the input data. The effect of the additional data included above to avoid the catastrophic behavior during MD runs is visible for short N-O distances in Fig.~\ref{fig:bond_stretching}(bottom). This data is more scarce than the original data centered around $d_\text{N-O}\approx$1.28~\AA. The increased uncertainty is however only reflected in the deep ensembles.

\subsection{A potential for bulk cubic \ce{SrTiO3}}
\begin{figure}
    \centering
    \includegraphics[width=.6\columnwidth]{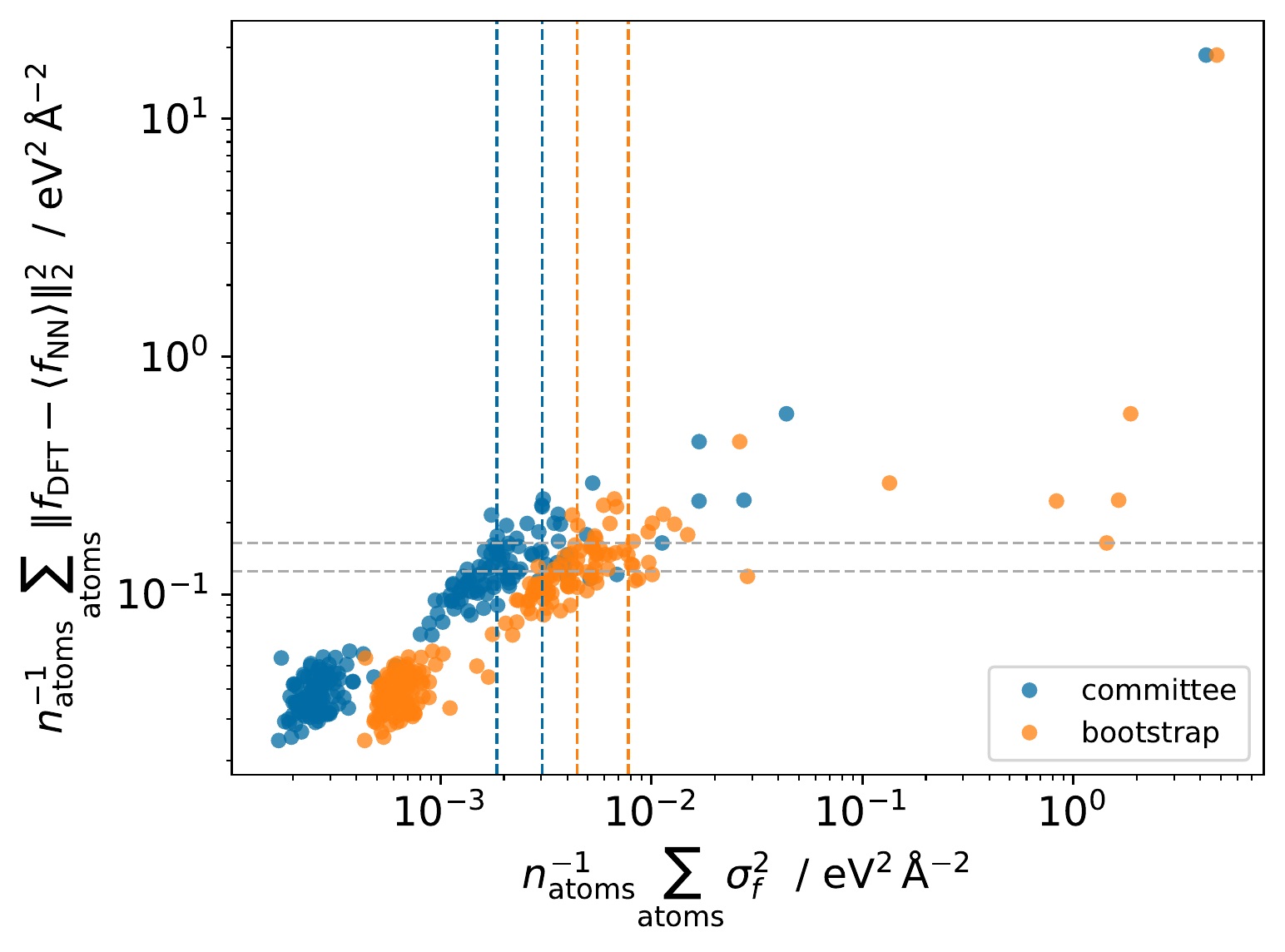}
    \caption{Comparison of error and uncertainty in the forces on the model trained on \ce{SrTiO3} bulk data. The uncertainties are calculated using either a bootstrap-aggregation ensemble or an NN committee. The points represent validation data and the lines are placed at the 75th and 90th percentiles of each quantity.}
    \label{fig:committee_training}
\end{figure}

The test statistics of the potential trained only on data for bulk \ce{SrTiO3} (a committee of $10$ NNs) are summarized by a $\mathrm{MAE}=\SI{2.44}{\milli\electronvolt\per\atom}$ and an $\mathrm{RMSE}=\SI{33.6}{\electronvolt\per\atom}$ for the energies, along with a $\mathrm{MAE}=\SI{115.91}{\milli\electronvolt\per\angstrom}$ and an $\mathrm{RMSE}=\SI{246.86}{\milli\electronvolt\per\angstrom}$ for the forces. The latter are to be judged in the context of  a mean absolute deviation of \SI{2.23}{\electronvolt\per\angstrom} and a standard deviation of \SI{4.53}{\electronvolt\per\angstrom} in the test set. For a bootstrap-aggregation ensemble with $10$ members, these same test statistics are $\mathrm{MAE}=\SI{2.54}{\milli\electronvolt\per\atom}$ and $\mathrm{RMSE}=\SI{38.18}{\electronvolt\per\atom}$ for the energies, and $\mathrm{MAE}=\SI{124.53}{\milli\electronvolt\per\angstrom}$ and $\mathrm{RMSE}=\SI{399.15}{\milli\electronvolt\per\angstrom}$ for the forces. A detailed comparison between the ground truth and the predictions of this potential can be seen in the top panel of Fig.~\ref{fig:committee_training}. That figure illustrates how well the committee estimate of the uncertainty works as a proxy for error and compares this performance with that of a bootstrap-aggregation ensemble. Although the sources of randomness in the committee are more limited (coefficient initialization only, without any additional diversity in the training/validation or the minibatch splits), and although the absolute values of their uncertainty estimates are different, both strategies succeed in identifying the most problematic points for the MLFF. Moreover, the Spearman correlation coefficient between uncertainty and error over the validation data set is 0.90 for the committee and 0.91 for the bootstrap-aggregation ensemble. Given the significant advantage in computational performance and scalability afforded by the committee, we settle on this approach for our treatment of \ce{SrTiO3}.

\begin{figure*}
    \centering
    \includegraphics[width=\textwidth]{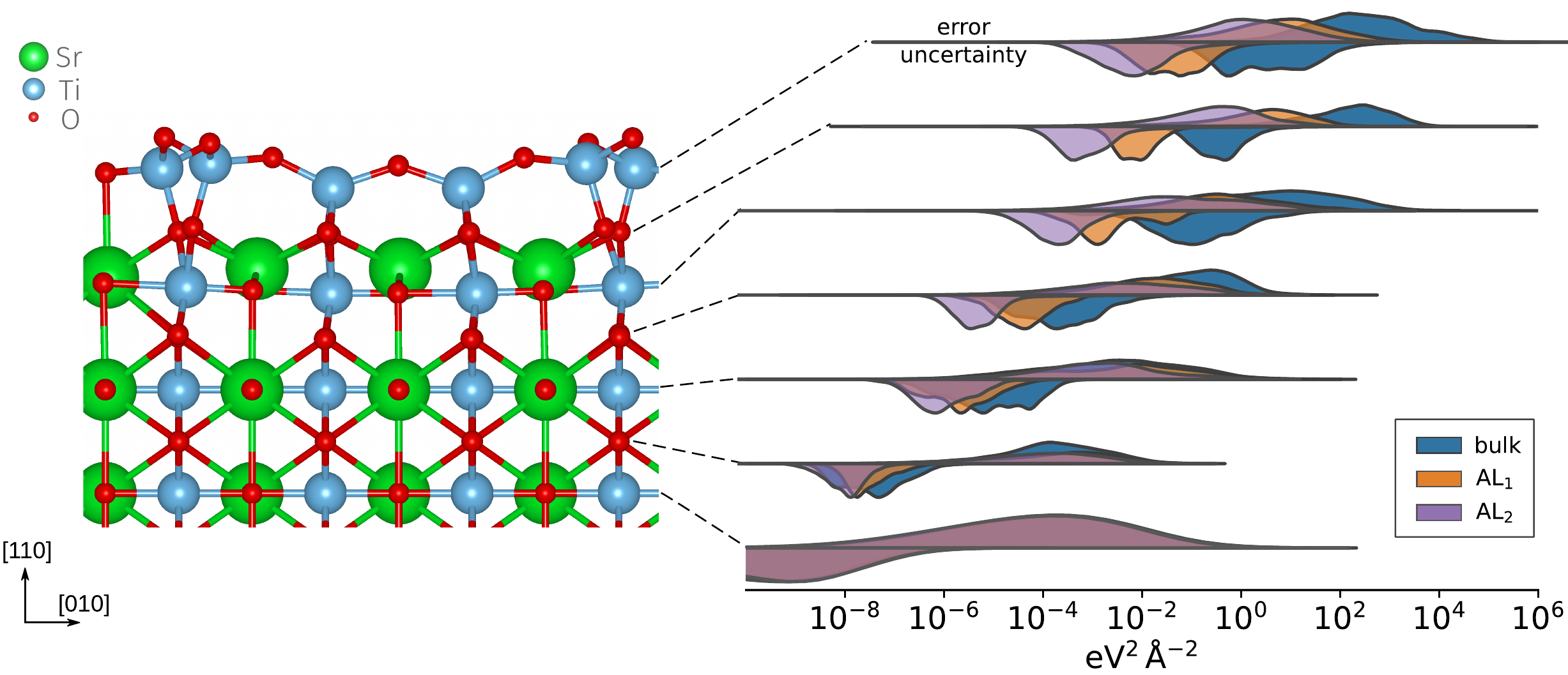}
    \caption{Split violin plot (right) showing the frequency density of the logarithm of the error (top) and the logarithm of the uncertainty (bottom) resolved by layer for the \ce{SrTiO3} bulk potential evaluated on the $4\times 1$ surface data and for the first and second refinements on that potential based on active learning. A side view of an \ce{SrTiO3}(110) $4\times 1$ slab is shown on the left, indicating the layers.}
    \label{fig:violins}
\end{figure*}

\begin{figure}
    \centering
    \includegraphics[width=\columnwidth]{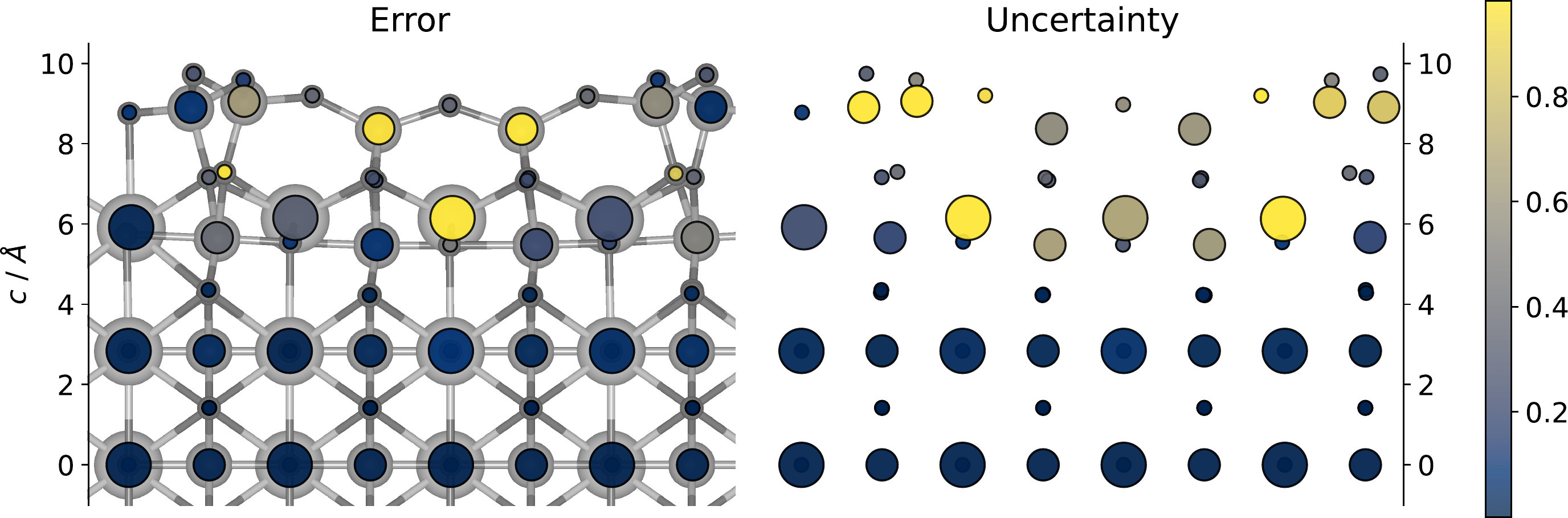}
    \caption{Details of the uncertainty estimate for the atomic forces of the \ce{SrTiO3} bulk potential when evaluated at the global minimum found in a structure search for $4\times 1$ surface data.\cite{STO_surfaces} The error is shown overlaid on a ball-and-stick model background as a guide to the eye.}
    \label{fig:color_bubbles}
\end{figure}

We now put this potential to the test on data for the $4\times 1$ surface reconstructions. Given the vastly greater variety of atomic environments found in those configurations, it is expected to perform poorly. Moreover, predictions of the force on atoms closer to the center of the slab, with more bulk-like neighborhoods and smaller displacements, can reasonably be expected to have lower associated errors. This should also be reflected in our uncertainty metric. Figures~\ref{fig:violins}~and~\ref{fig:color_bubbles} illustrate that this is indeed the case. Specifically, Fig.~\ref{fig:violins} shows how typical values of both the error and the uncertainty increase by orders of magnitude for atoms close to the free surface. This kind of spatially resolved analysis is only made possible by our reliance on forces instead of the aggregate energy. Note, however, that the correlation between uncertainty and error within the topmost layers is not good enough to identify the most problematic atoms directly (see Fig.~\ref{fig:color_bubbles}). This shows that the skill of the uncertainty in the forces as a proxy for error improves with the cardinal of the set of atoms considered: it is still good when applied to a whole layer, but less so for individual atoms. Another factor influencing this result is that, although the contribution of an atom to the energy depends only on its own descriptors, the force on that atom depends on the descriptors of all neighbors in its local environment. Therefore, atoms with neighbors showing high uncertainties in their forces should also be considered candidates for high errors. As a consequence, simply looking for underrepresented types of local environments is not enough to improve the quality of a training set when tackling a new system such as these $4\times 1$ reconstructions.

\begin{figure}
    \centering
    \includegraphics[width=.6\columnwidth]{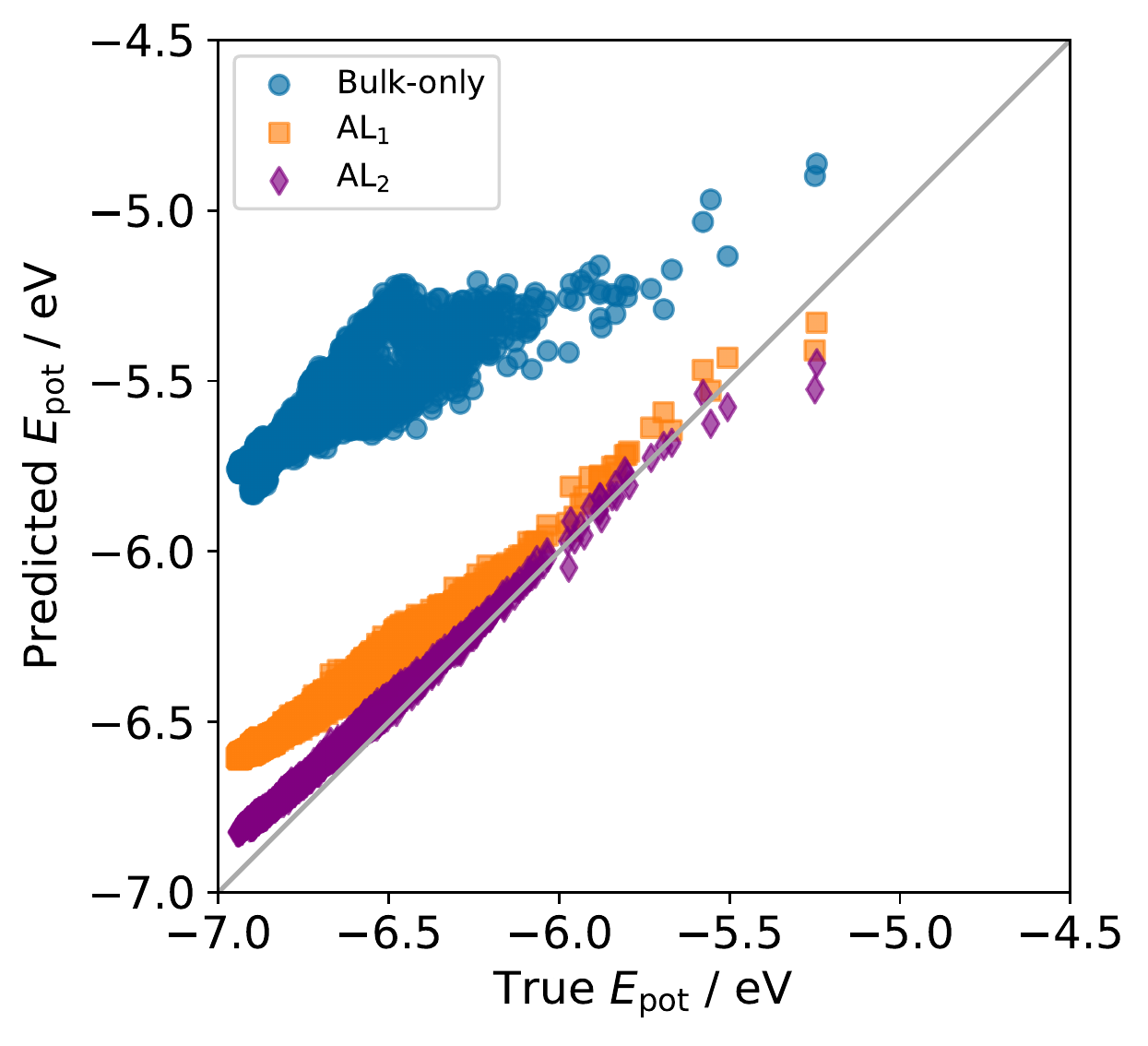}
    \caption{Parity plot for the energies of the test set of the $4\times 1$ \ce{SrTiO3} surface using three different potentials in an active-learning chain: one trained using only bulk data, one (AL$_1$) including $100$ additional structures generated through the maximization of the adversarial loss computed from the bulk-only potential and one (AL$_{2}$) whose training data set contains a further $100$ structures created through the same procedure but based on AL$_1$.}
    \label{fig:AL_energies}
\end{figure}

The accuracy delivered by the bulk-trained NNFF when applied to the surface atoms is clearly insufficient even for applications requiring only a rough approximation to the forces. Those huge errors are also enough to completely distort the global picture of the potential energy hypersurface: as the blue series in Fig.~\ref{fig:AL_energies} illustrates, not only is the correlation between the predicted and true energies rather poor, but there is a systematic error that can be interpreted as a misalignment between the origins of energies of the DFT calculations and the FF. 

\subsection{Actively learning a potential for the \ce{SrTiO3} surface}
Augmenting the training data set with relevant configurations can be expected to alleviate this situation. To assess the extent to which this is the case, we train a new potential, AL$_1$, after adding $100$ configurations generated through the adversarial algorithm described above. The effect on the potential energy is dramatic, as shown by the orange series in Fig.~\ref{fig:AL_energies}: correlation is greatly improved (as attested by the more elongated form of the graph) and the problematic offset mostly disappears. However, the energies predicted for the lowest-lying configurations still stand out as far worse than those for less stable configurations. Given the importance of surface reconstruction in determining the ground state and the fact that the bias still trends in the same direction as the offset of the bulk-only potential, this is likely to be due to the surface layers. The second series of violin plots for the forces in Fig.~\ref{fig:violins} confirms this hypothesis: both the big improvements in predictive skill and the remaining inaccuracies can be predominantly traced to the surface layers.

A second round of active learning using an identical process starting from the same single configuration yields diminishing returns. However, if each iteration of the new adversarial process to generate an additional set of training $100$ configurations is instead started from one of the $100$ created for the first round, chosen at random, we achieve another major improvement in the uncertainty and error of the forces, as shown by the violet violins in Fig.~\ref{fig:violins}. After that second round, the uncertainties in adjacent layers become much more comparable. A notable feature of this progression is that the committee uncertainty estimate decreases more than the actual error. This suggests that the kind of uncertainty estimates analyzed here are more useful as a tool for comparing different points evaluated using the same potential, and less reliable when it comes to comparing different potentials. The total potential energy predictions also get visibly closer to parity with DFT (see Fig.~\ref{fig:AL_energies}), especially for the most difficult configurations.

The importance of starting each round of active learning from  as diverse a set of configurations as possible is best understood in light of the origin of our test data for \ce{SrTiO3}, which comes from an evolutionary search using the CMA-ES algorithm. The central consideration in such a search is balancing exploration and exploitation of configuration space, and as such they try not to stay in the local environment of a single local minimum. This stands in stark contrast with the local optimization of the adversarial loss at the center of our active-learning procedure (as prescribed in Ref.~\onlinecite{Bombarelli_adversarial}). 

\section{Summary and conclusions}

We implemented three different low-overhead uncertainty estimators on top of an automatically differentiable descriptor-based NNFF and tested them in two different applications of practical relevance designed to involve excursions outside of the regions covered by training data. Two of those estimators are based on deep ensembles of NNs with additional heads that provide an intrinsic assessment of the difficulty of a prediction, while the third is the variance of a simple committee of models.

We first ran MD simulations of EAN using a potential trained on DFT forces and energies for snapshots of an MD trajectory generated with a different FF. All the uncertainty metrics, whether applied to forces or energies, are able to detect when the potential leaves its comfort zone and predict catastrophic failures of the MD simulation. Moreover, they provide a viable criterion for selecting configurations from which useful information can be harvested for retraining the FF. After applying a single iteration of this procedure to snapshots of the problematic trajectories we were able to rerun the trajectories without issues. The same estimators show skill similar to a bootstrap-aggregation ensemble but can be trained in a small fraction of the time.

In our second example we took the first steps towards learning a potential to describe the $4\times 1$ surface reconstructions of \ce{SrTiO3} starting from a potential developed for the bulk solid. We followed an adversarial learning approach initially designed for molecules; our results illustrate both its efficiency and the limitations its local character entails when used in the context of a global structure search. Key to enabling this active-learning workflow in practice is a dramatically accelerated training cycle brought about by two innovations in our NNFF design: residual learning and a nonlinear learned optimizer replacing the usual stochastic gradient descent schemes.

Collectively, the results of these two examples underscore the amount of information contained in the forces and their uncertainties when compared to the total energy of the system, in keeping with existing observations about the advantages of forces over energies in training. When evaluated for the same quantity (energies or forces), heteroscedastic models and the associated uncertainty metrics do not seem to offer a systematic advantages over a simple committee in terms of identification of difficult or uncertain configurations, but can provide more specific information about the kind of uncertainty being estimated (aleatoric or epistemic).

\begin{acknowledgments}
This work was supported by the Austrian Science Fund (FWF) (SFB F81 TACO). E.H. acknowledges support from the Austrian Science Fund (FWF), project J-4415. The financial support of the Spanish Ministry of Science and Innovation (PID2021-126148NA-I00 funded by MCIN/AEI/10.13039/501100011033/FEDER, UE) are gratefully acknowledged. H.M.C. thanks the USC for his \enquote{Convocatoria de Recualificación do Sistema Universitario Español-Margarita Salas} postdoctoral grant under the \enquote{Plan de Recuperación Transformación} program funded by the Spanish Ministry of Universities with European Union's NextGenerationEU funds. This work was supported by the Fundacão para a Ciência e Tecnologia (FCT) (funded by national funds through the FCT/MCTES (PIDDAC)) to CIQUP, Faculty of Science, University of Porto (Project UIDB/00081/2020), IMS-Institute of Molecular Sciences (LA/P/0056/2020).
\end{acknowledgments}

\appendix

\section{Obtaining a variance of the forces from statistics of the energies}\label{apx:covariance}
Let $f_i\supar{\alpha}=-\diffp{E_{\mathrm{pot}}}{X_i\supar{\alpha}}$ be the component along Cartesian axis $\alpha$ of the force on atom $i$. We denote by $\aqty*{Q}$ the expected value of a quantity $Q$ according to any relevant distribution (average over members of an ensemble, mathematical expectation over all possible initial weights and biases\ldots). The uncertainty in $f_i\supar{\alpha}$ can be quantified through its standard deviation $\sigma_{f_i\supar{\alpha}}$, the square root of its variance $\sigma^2_{f_i\supar{\alpha}}$, which is, in turn, the special case with $i=j$ and $\alpha=\beta$ of the covariance between two components of the force,

\begin{equation}
    \cov\sqty*{f_i\supar{\alpha}, f_j\supar{\beta}} = \aqty*{\sqty*{f_i\supar{\alpha} - \aqty*{f_i\supar{\alpha}}}\sqty*{f_j\supar{\beta} - \aqty*{f_j\supar{\beta}}}}.
    \label{eqn:covariance}
\end{equation}

\noindent This can be more compactly written as $\cov\sqty*{f_i\supar{\alpha}, f_j\supar{\beta}}=\aqty*{\delta f_i\supar{\alpha}\delta f_j\supar{\alpha}}$ if we introduce the shorthand notation $\delta Q = Q - \aqty*{Q}$. Explicitly expressing the derivatives in the definition of the forces as limits, we get

\begin{equation*}
    \begin{aligned}
        \cov\sqty*{f_i\supar{\alpha}, f_j\supar{\beta}} =
        \Biggl \langle &
        \sqty*{
            \lim\limits_{h\longrightarrow 0} \frac{\delta E_{\mathrm{pot}}\bqty*{\bm{X} + h \hat{\bm{u}}_i\supar{\alpha}}-\delta E_{\mathrm{pot}}\pqty*{\bm{X}}}{h}
        }                \\
        &\sqty*{
            \lim\limits_{H\longrightarrow 0} \frac{\delta E_{\mathrm{pot}}\bqty*{\bm{X} + H \hat{\bm{u}}_j\supar{\beta}}-\delta E_{\mathrm{pot}}\pqty*{\bm{X}}}{H}
        }
        \Biggr \rangle,
    \end{aligned}
\end{equation*}

\noindent where $\hat{\bm{u}}_i\supar{\alpha}$ is the unit vector $\diffp{\bm{X}}{X_i\supar{\alpha}}$. Expanding the product and using the linear character of the expected values, the above can be recast as

\begin{equation}
    \begin{aligned}
    \cov\sqty*{f_i\supar{\alpha}, f_j\supar{\beta}} = \lim\limits_{H,h\longrightarrow 0}\frac{1}{hH}
    \Biggl\lbrace &
    \aqty*{\delta E_{\mathrm{pot}}\bqty*{\bm{X} + h \hat{\bm{u}}_i\supar{\alpha}}\delta E_{\mathrm{pot}}\bqty*{\bm{X} + H \hat{\bm{u}}_j\supar{\beta}}}
    + \aqty*{\sqty*{\delta E_{\mathrm{pot}}\bqty*{\bm{X}}}^2}\\
    -&\aqty*{\delta E_{\mathrm{pot}}\bqty*{\bm{X} + h \hat{\bm{u}}_i\supar{\alpha}}\delta E_{\mathrm{pot}}\bqty*{\bm{X}}} - \aqty*{\delta E_{\mathrm{pot}}\bqty*{\bm{X}}\delta E_{\mathrm{pot}}\bqty*{\bm{X} + H \hat{\bm{u}}_j\supar{\beta}}}
    \Biggr\rbrace.
    \end{aligned}
    \label{eqn:expanded_covariance}
\end{equation}

\noindent If the limit exists, it can be calculated along the line $h=H$, which turns Eq.~\eqref{eqn:expanded_covariance} into the two-point formula for the mixed second derivative

\begin{equation*}
    \cov\sqty*{f_i\supar{\alpha}, f_j\supar{\beta}} = \diffp{\aqty*{\delta E_{\mathrm{pot}}\pqty*{\bm{X}}\delta E_{\mathrm{pot}}\pqty*{\bm{Y}}}}{X_i\supar{\alpha},Y_j\supar{\beta}}[\bm{X}=\bm{Y}]
\end{equation*}

\noindent In particular,

\begin{equation}
    \sigma^2_{f_i\supar{\alpha}}=\diffp{\aqty*{\delta E_{\mathrm{pot}}\pqty*{\bm{X}}\delta E_{\mathrm{pot}}\pqty*{\bm{Y}}}}{X_i\supar{\alpha},Y_i\supar{\alpha}}[\bm{X}=\bm{Y}].
\label{eqn:covariance_kernel}
\end{equation}

Thus, having a differentiable estimate of the uncertainty of $E_{\mathrm{pot}}$ at a single point is not enough to obtain an estimate of the uncertainty of $f_i\supar{\alpha}$. The latter requires being able to calculate the covariance between the predictions of $E_{\mathrm{pot}}$ for two different points of configuration space in close proximity. The Behler-Parrinello architecture and other set-pooling architectures are designed to provide predictions for a single point $\bm{X}$ and therefore ill-suited for the addition of a new head that can output an intrinsic estimate of the spatial covariance $\aqty*{\delta E_{\mathrm{pot}}\pqty*{\bm{X}}\delta E_{\mathrm{pot}}\pqty*{\bm{Y}}}$. However, Eq.~\eqref{eqn:covariance_kernel} may prove a good match for alternative approaches like the recently proposed pairwise difference regression\cite{padre} (PADRE) that explicitly take two configurations as inputs and can operate with statistics over such pairs.

\bibliography{uncertainties}

\end{document}